\documentclass[twocolumn,aps,prl,superscriptaddress,showpacs]{revtex4}
\usepackage{amsmath,bm}%
\usepackage{graphicx}%

\begin{document}

\draft
\title{ A proposed reaction channel for the synthesis of the superheavy nucleus Z = 109}
\thanks{ Supported by the Major State Basic Research Development Program under Contract No G200077400, the
National Natural Science Foundation of China under (NNSFC) Grant
No 10135030, and the CAS Grant for
 Distinguished Young Scholars of NNSFC under Grant No 19725521.}

\author{K. Wang} 
\affiliation{Shanghai Institute of Applied Physics, Chinese
Academy of Sciences, P. O. Box 800-204, Shanghai 201800}
\author{Y. G. Ma} \thanks{To whom correspondence should be addressed. Email: ygma@sinr.ac.cn}
  \affiliation{Shanghai Institute of  Applied Physics, Chinese Academy of Sciences, P. O. Box 800-204,
Shanghai 201800}
   \affiliation{   CCAST (World Laboratory), P. O. Box 8730, Beijing 100080}
\author{G. L. Ma}
 \author{ Y. B. Wei}
\author{X. Z. Cai}
\author{J. G. Chen}
\author{ W. Guo}
\affiliation{Shanghai Institute of  Applied Physics, Chinese
Academy of Sciences, P. O. Box 800-204, Shanghai 201800}
\author{ C. Zhong}
\affiliation{Shanghai Institute of  Applied Physics, Chinese
Academy of Sciences, P. O. Box 800-204, Shanghai 201800}
\author{ W. Q. Shen}
  \affiliation{Shanghai Institute of  Applied Physics, Chinese Academy of Sciences, P. O. Box 800-204,
Shanghai 201800} \affiliation{   CCAST (World Laboratory), P. O.
Box 8730, Beijing 100080}
  \affiliation{  Department of Physics, Ningbo University, Ningbo 315211}
\date{Received 14 Nov 2003; Pubished in Chin. Phys. Lett. 21, 464 (2004) }

\begin{abstract}
We apply  a statistical-evaporation model (HIVAP) to calculate the
cross sections of superheavy elements, mainly about actinide
targets and compare with some available experimental data.  A
reaction channel $^{30}$Si + $^{243}$Am
is proposed  for the synthesis of the element Z = 109 and the
cross section  is estimated.
\end{abstract}


\maketitle

The synthesis of superheavy elements always attract the attentions of the
nuclear physicists and chemists  since the superheavy
island was predicted in 1960. Now  it has been taken as one of the major
research directions in main laboratories of nuclear physics around the world.
So far, the heaviest elements  Z = 114 and 116 were synthesized and identified
 in Dubna recently \cite{114,116}. Some structure and dynamical properties of heavy or superheavy
nuclei have been investigated in various aspects
\cite{Ren,Bao,Wang,Li,Meng,Zhao,Ma}. However, theoretical efforts
on the production cross section of superheavy elements are being
waited for further improvement since most theoretical models are
not able to predict the credible cross sections yet to provide the
valid suggestions for the planning of experiments, hence the
experiments have been performed more or less in  the empirical
ways.

Nevertheless,  it is still interesting to do some semi-empirical calculations
to explore a possible way to help the experimental design.
 To this end, firstly, we perform the systematical calculations and
let them agree with the experimental data well. Based on the good
reproduction to some known data, we can make a reasonable
extrapolation for the production cross section of some superheavy
nuclei. In this paper we would like to explore this possibility
with help of a statistical-evaporation model, so-called HIVAP, so
that we can make a believable suggestion for the experimental
proposal in our national laboratory of Heavy Ion Research Facility
in Lanzhou (HIRFL).

HIVAP is a statistical-evaporation model
 \cite{Reisdorf,Reisdorf2,martin1,martin2,Reisdorf3},
which assumed that the process of synthesizing superheavy nuclei
includes two stages:  firstly the projectile and target nuclei
completely fuse to a compound nucleus, then the compound nucleus
de-excites by fission or emitting light particles and
$\gamma$-rays.  The complete cross section $\sigma(E_{cm})$ is
assumed as
\begin{equation}
\sigma(E_{cm}) = \pi \lambda^2 \sum^{J_{max}}_{J=0}
(2J+1)T(J,E_{cm})P(J,E^*)
\end{equation}
where $\lambda$ is the de Broglie wave length of relative motion
of the colliding nuclei, $E_{cm}$ is the energy of center-of-mass,
$E^*$ is the excitation energy. $J$ is the total angular momentum
quantum number, the upper limit $J_{max}$ is obtained by
\cite{CPS}. $T(J,E_{cm})$ is the fusion probabilities of the $J$th
partial wave through the Coulomb barrier. $P(J,E_{cm})$ is the
survival probability of the residue evaporation nucleus after the
compound nucleus de-excites by fission or emitting light particles
passes the Coulomb barrier and is captured.

 The fusion probabilities $T(J,E_{cm})$ is assumed by \cite{pt} as
\begin{equation}
T(J,E_{cm}) = \sum_{V'_{B}} f(V'_B) t_J(E_{cm},V'_B)
\end{equation}
where $f(V'_B)$ is a quasi-gaussian fusion barrier
distribution, $t_J$ is the transmission coefficient obtained by WKB
approximation.
 The fusion potential\cite{Reisdorf3} is given by
\begin{equation}
V_{R} = V_{coul}-V_{0}  exp[1.33(c_{p} + c_{t}-R)/b] c
\end{equation}
where $V_{Coul}$ is the Coulomb potential, $c_{p}$ and $c_{t}$ are
the central radii \cite{radii} of projectile and target,
c = $c_{p}$$c_{t}$/($c_{p}$+$c_{t}$),  b is the surface diffuseness
parameter \cite{radii}.
 The $V'_B$ in eq.(2) is average barrier which was adjusted by a fixed
 constant $V_0$ (in MeV/fm) \cite{V0} in eq.(3). The fluctuation around
 $V'_B$ is scaled by a parameter $r_0$.
 We have adopted a Gaussian distribution
 cut  off at both ends at $r_0 \pm t \sigma_B (r_0)$ , t is a cut-off parameter \cite{pt}.
We use t = 3 in our calculation.
  $\sigma_B(r_0)$ is standard deviation of $r_0$ distribution.

 The level density is an important factor,  the progress of
synthesizing superheavy nucleus can be explained by the
competition of fusion with a fast fission-like process which can
be identified with quasi-fission \cite{Shen}. Then the survival
probability of residue evaporation nucleus can be expressed using
the level densities in the compound and equilibrium configurations
as \cite{martin3}:
\begin{equation}
P( J,E^*)= \frac{\rho(J,E^*_{cn})}{\rho(J,E^*_{cn}) +
\rho(J,E^*_{eq})}
\end{equation}
where $\rho(J,E^*_{cn})$ is the level density of compound
configuration and $\rho(J,E^*_{eq})$  is the level density of
equilibrium configuration. The level density is
 \begin{eqnarray}
\rho(J,E^*) & = & \frac{1}{24}(\frac{\hbar}{2\theta})^{3/2}(2J+1) a^{1/2}U^{-2}_J exp[2(aU_J)^{1/2}] \nonumber, \\
U_J& = & E^*-E_r(J) + P_{pair}
\end{eqnarray}
 $P_{pair}$ is the pairing correction obtained from experimental odd-even mass fluctuations.
  $a$ is level density parameter, obtained from
   \begin{equation}
            a= \overline{a} [1 + f(E^*)B_f^{shell}/E^*],
\end{equation}
and
   \begin{equation}
   f(E^*)=1-exp(-E^*/E_d),
\end{equation}
  in which $E_d$ is the damping energy, and
   \begin{equation}
   \overline{a} = 0.04543r_0^3A + 0.1355r_0^2A^{2/3} +
   0.1426r_0A^{1/3},
\end{equation}
where $E_\gamma (J)$ is the yrast energy of either the equilibrium
configuration (light-particle and $\gamma$-emission) or the
saddle-point configuration (fission), it reads
   \begin{equation}
　　　　　E_\gamma (J) = J(J+1)\hbar^2/2\theta,
\end{equation}
in which  $\theta$ is the moment of inertia.  The fission barrier
is  defined by including the liquid drop component ($B_f^{LD}$)
 and the shell component ($B_f^{shell}$),  scaled by a coefficient $C$, i.e,
 \begin{equation}
    B_f = C (B_f^{LD} +  B_f^{shell}),
\end{equation}
 in equilibrium configuration.

HIVAP takes into account the competition of $\gamma$-ray, neutron,
proton and $\alpha$-particle emission with fission using
angular-momentum and shape-dependent level densities and
angular-momentum-dependent fission barriers.


The level densities have been calculated using the well known
Fermi gas model. The same level density parameters have been used
for the fission and neutron emission channels ($a_f$/$a_n$ =
1). The arguments in support of this value were discussed in
Ref. \cite{afav}.  A phenomenological way of the introduction of
shell effects into the level density calculation according to
Ignatyuk \cite{Ignatyuk}, have been used in the evaporation
channels. The shell damping energy $E_d$ is 18.5 MeV. The
liquid-drop fission barrier ($B_f^{LD}$)  has been calculated
according to the rotating charged liquid drop model of
Cohen-Plasil-Swiatecki \cite{CPS}. The shell component
($B_f^{shell}$) of the fission barrier is equal to the
difference between the liquid-drop model \cite{liquid} and
experimental \cite{mass} masses of the nucleus. Light-particle
transmission coefficients are obtained using WKB approximation.
The shell corrections are added at the saddle point.

In Fig.1  we use HIVAP model to calculate the fusion cross sections for the
 reaction of $^{18}$O + $^{248}$Cm. The cross sections for the  channels
with 3n to 7n emission are shown. Generally the energy corresponding to the
peak of the cross section for a given xn channel increases with the number
of emitted neutrons, which is consistent with the excitation extent of the
reaction with the more neutron emission. In all channels, the maximum
cross section can be reached for 5n channels.
  Since $^{18}$O + $^{248}$Cm is a hot fusion reaction, we take the
maximum value of cross section at 5n channel to compare with the
experimental data \cite{data0,data1,data3}. Fig. 2 shows the
comparison. An overall satisfied overlap of the calculations with
the data was obtained.  We listed the experiment data in Table I.
From the comparison, it looks that this model has a good agreement
with the data for these reactions.

\begin{figure}
\vspace{-0.6truein}
\includegraphics[scale=0.4]{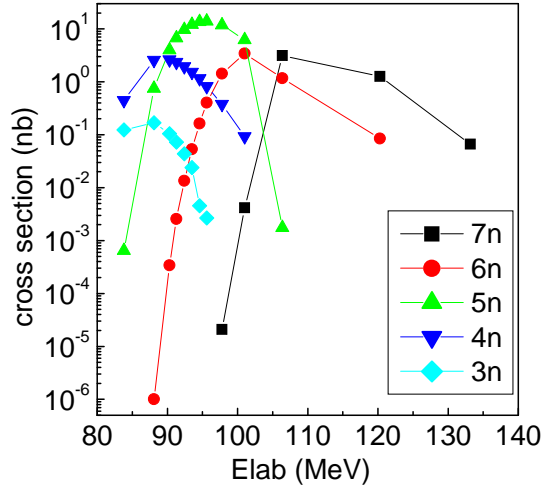}
\vspace{-0.4truein}
\caption{\footnotesize Calculated evaporation residue cross sections
for $^{18}O+^{248}Cm$ in different xn channels. }
\label{fig1}
\end{figure}

\begin{figure}
\vspace{-0.6truein}
\includegraphics[scale=0.4]{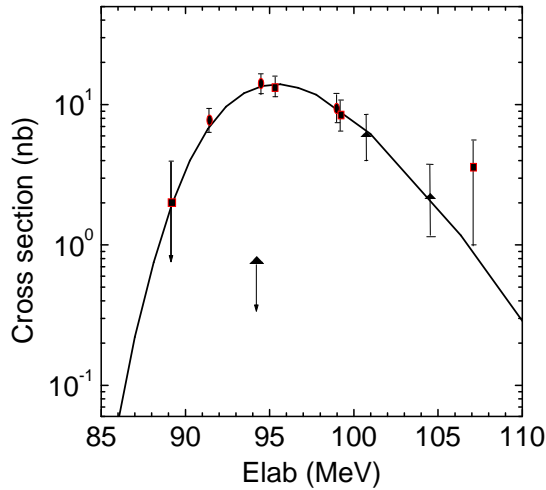}
\vspace{-0.4truein} \caption{\footnotesize Comparison the
calculation results by HIVAP (line) with the experimental data
(symbols). The different symbols present  the results from the
different works which have been already listed in Table I.}
\label{fig2}
\end{figure}

So far, we used the HIVAP to fit the experimental data very well.
Based on this achievement, we would like to make some predictions
for the synthesizing the  elements of Z =109. With all the same
parameters in the model calculation, we  calculated some channels
for producing Z =109 elements and listed the results is in Table
II. From this table, we found that the channels
 of $^{30}Si + ^{243}Am \rightarrow ^{270}Mt + 3n$ at E = 151 MeV or
$^{30}Si+ ^{243}Am \rightarrow ^{269} Mt + 4n$ at E=161 MeV have a
larger cross sections for  synthesizing the  new isotopes of
element 109.  The cross section can reach to $\simeq$ 21 pb.

\begin{table}
\caption{\label{tab:table3} The calculation of cross sections for
synthesizing Z = 103-107 calculate by HIVAP. "*" represent the
 reaction which has been finished in LanZhou (HIRFL) recently
but no data published yet. The result which calculated by Alice code
is 1.8nb \cite{Alice}.   }
\begin{ruledtabular}
\begin{tabular}{ccccc}
channel & $E_{Lab}$ & $\sigma_{exp}$  &  Reference & HIVAP\\
 & (MeV) & (nb) &  & (nb)\\
  \colrule
$^{248}$Cm($^{18}$O,5n)$^{261}$Rf & 94.2 & $<$ 0.9 & \cite{data3} \\
$^{248}$Cm($^{18}$O,5n)$^{261}$Rf & 97 & 5 & \cite{data1} & 13\\
$^{248}$Cm($^{18}$O,5n)$^{261}$Rf & 100.4 & $4.5^{+3.5}_{-2.0}$ & \cite{data3} & 12\\
$^{248}$Cm($^{18}$O,5n)$^{261}$Rf & 103.9 & $1.7^{+3.3}_{-1.0}$ & \cite{data3} & 4.4\\
  \colrule
$^{248}$Cm($^{18}$O,5n)$^{261}$Rf & 91 & 8 $\pm$  2& \cite{data0} & 4.2\\
$^{248}$Cm($^{18}$O,5n)$^{261}$Rf & 94 & 13 $\pm$ 3& \cite{data0} & 11.3\\
$^{248}$Cm($^{18}$O,5n)$^{261}$Rf  &  99  & 8 $\pm$ 2 & \cite{data0} & 15\\
  \colrule
$^{244}$Pu($^{22}$Ne,5n)$^{261}$Rf & 114 & 4.4 & \cite{data16} & 7.8\\
$^{244}$Pu($^{22}$Ne,5n)$^{261}$Rf & 120 & 3.8 & \cite{data16} & 4.6\\
  \colrule
$^{248}$Cm($^{19}$F,5n)$^{262}$Db & 106 & 2 & \cite{data6}\\
$^{248}$Cm($^{19}$F,5n)$^{262}$Db & 106 & 1.3 $\pm$ 0.4 & \cite{data0} & 0.67\\
$^{248}$Cm($^{19}$F,5n)$^{262}$Db & 106.5 & $0.26^{+0.15}_{-0.09}$ & \cite{data5} \\
  \colrule
$^{248}$Cm($^{15}$N,5n)$^{258}$Lr & 88 & 200 & \cite{dataCm} & 480 \\
  \colrule
$^{248}$Cm($^{15}$N,4n)$^{259}$Lr & 80 & 40 &  \cite{dataCm} & 169\\
  \colrule
$^{249}$Cf($^{18}$O,4n)$^{263}$Sg & 95 & 1 & \cite{dataCfo} & 2.3\\
  \colrule
$^{249}$Cf($^{15}$N,4n)$^{260}$Db & 84 & 3.5 & \cite{dataCfn} & 13\\
  \colrule
$^{243}$Am($^{18}$O,5n)$^{261}$Lr & 96 & 30 & \cite{dataAm} & 149\\
  \colrule
$^{249}$Bk($^{15}$N,4n)$^{260}$Rf & 82 & 18 & \cite{dataBk} & 31\\
  \colrule
$^{249}$Bk($^{18}$O,5n)$^{262}$Db & 99 & 6 $\pm$ 3 & \cite{data4} & 2.1\\
  \colrule
$^{241}$Am($^{22}$Ne,4n)$^{259}$Db & 118 & 1.6 $\pm$ 1.2 & \cite{db259} & 2.7\\
  \colrule
$^{243}$Am($^{26}$Mg,4n)$^{265}$Db & 131 & N/A & * & 0.097\\
\end{tabular}
\end{ruledtabular}
\end{table}


 From the calculation results listed in Table 1 we can see the maximum
error of our calculations is within  5 times comparing with the
experimental data, so it seems that our estimation for cross
section should be reasonable. Element 109 was produced firstly in
cold fusion by physicists of the Heavy Ion Research Laboratory,
Darmstadt, West Germany using $^{58}$Fe ion bombarding on
$^{209}$Bi target. For element 109, the isotopes of $^{268}Mt$ and
$^{266}Mt$ were already synthesized in GSI \cite{gsi109}. We
suggest that the channel of $^{30}Si + ^{243}Am \rightarrow
^{270}Mt + 3n$ or $^{30}Si+ ^{243}Am \rightarrow ^{269}Mt + 4n$
to synthesize new isotopes of element 109.  $^{30}$Si is a
suitable projectile since there is 3.1$\%$ $^{30}$Si contained in
element Si and it is a stable nucleus.  Considering the present
situations of our national laboratory in Lanzhou (HIRFL), it is
very difficult to synthesize the very heavy elements such as above
Z = 112 etc. in the moment, but it is feasible to try to search
for the element 109 with the proposed reaction channel.

\begin{table}
\caption{\label{tab:table4}The calculated cross section for
synthesizing element 109. }
\begin{ruledtabular}
\begin{tabular}{ccc}
 Channel & $E_{lab}$(MeV) & Cross section (pb) \\ \hline
 $^{254}$Es ($^{22}$Ne,4n) $^{272}$Mt & 115 & 8    \\
 \colrule
$^{254}$Es ($^{22}$Ne,5n) $^{271}$Mt & 124 & 5    \\
 \colrule
$^{254}$Es ($^{20}$Ne,4n) $^{270}$Mt & 114 & 3 \\
 \colrule
$^{254}$Es ($^{20}$Ne,5n) $^{269}$Mt & 120& 3    \\
 \colrule
$^{249}$Bk ($^{26}$Mg,4n) $^{271}$Mt & 137 & 9.5\\
 \colrule
$^{249}$Bk ($^{26}$Mg,5n) $^{270}$Mt & 148 & 5 \\
 \colrule
$^{243}$Am ($^{28}$Si,4n) $^{267}$Mt & 155 & 3\\
 \colrule
$^{243}$Am ($^{30}$Si,4n) $^{269}$Mt & 161.3 & 13\\
 \colrule
$^{243}$Am ($^{30}$Si,3n) $^{270}$Mt &151 &21.6
\end{tabular}
\end{ruledtabular}
\end{table}

To recap, the extensive experimental data for the synthesis of
superheavy nuclei have been well reproduced in our calculations
with help of the HIVAP code. It illustrates that the
statistical-evaporation model can still work for  the production
of superheavy nuclei in a suitable mass region. Within a
reasonable extrapolation, we propose to use
 a new channel, i.e. $^{30}Si + ^{243}Am \rightarrow ^{270}Mt + 3n$
or $^{30}Si+ ^{243}Am \rightarrow ^{269} Mt + 4n$
to synthesize new isotopes of element 109.

Ma Y G would like to appreciate Dr. Martin Veselsky for providing HIVAP code
and worthy discussions.

{}
\end{document}